# Integrated Room Temperature Single Photon Source for Quantum Key Distribution

Helen Zhi Jie Zeng[1], Minh Anh Phan Ngyuen[1], Xiaoyu Ai[2], Adam Bennet[3], Alexander Solnstev[1], Arne Laucht[2], Ali Al-Juboori[2], Milos Toth[1,4], Rich Mildren[3], Robert Malaney[2], and Igor Aharonovich[1,4]

[1]*School of Mathematical and Physical Sciences, Faculty of Science, University of Technology Sydney, Ultimo, New South Wales, 2007, Australia*
[2]*School of Electrical Engineering and Telecommunications, The University of New South Wales, Sydney, NSW 2052, Australia*
[3]*MQ Photonics Research Centre, Department of Physics and Astronomy, Macquarie University, Sydney, NSW 2109, Australia*
[4]*ARC Centre of Excellence for Transformative Meta-Optical Systems (TMOS) University of Technology Sydney, Ultimo, New South Wales , Australia*
*Corresponding author: igor.aharonovich@uts.edu.au*

**High-purity single photon sources (SPS) that can operate at room temperature are highly desirable for a myriad of applications, including quantum photonics and quantum key distribution. In this work, we realise an ultra-bright solid-state SPS based on an atomic defect in hexagonal boron nitride (hBN) integrated with a solid immersion lens (SIL). The SIL increases the source efficiency by a factor of six, and the integrated system is capable of producing over ten million single photons per second at room temperature. Our results are promising for practical applications of SPS in quantum communication protocols.**

Single-photon sources (SPSs) are the cornerstone of quantum optics, offering a reliable way to deterministically generate high-purity photons on demand[1, 2]. A plethora of applications exist to utilise these sources ranging from quantum information processing and computing to quantum cryptography[3-6], including efficient implementation of quantum key distribution (QKD) protocols[6-8]. However, practical QKD requires several SPS attributes to be addressed collectively, including brightness, purity and stability. Hence, there is a clear demand for such a source to be engineered and packaged in an integrated photonic system.

Hexagonal boron nitride (hBN) is of particular interest in this space as a host of atomic defects that can be deployed as high quality SPSs, featuring both exceptional brightness, stability and good single-photon purity (the probability that there is no more than one photon per pulse)[9-15]. In contrast to quantum-dot based counterparts that require cryogenic cooling[1], SPSs based on hBN operate at room temperature (RT), offering a practical advantage for applications in quantum communication. However, the main drawback of all solid-state SPSs, is a limited excitation efficiency and/or collection efficiency due to trapping of light in the host crystal. There have been multiple works aiming to improve SPS performance by increasing both the internal quantum efficiency[16-18] and collection efficiency[19, 20]. However, most approaches rely on precise emitter positioning and/or nano-fabrication, making them complex, difficult to scale and unsuitable for mass-production.

In this work, we develop and realise an integrated SPS based on hBN and a solid immersion lens (SILs)[21-23]. This approach is promising as SILs are easy to fabricate and available commercially. We show that examples of the integrated hBN - SIL device offer over a six-fold increase in photon collection efficiency, yielding a single-photon collection rate of $10^7$ Hz, and are also able to maintain an excellent purity of $g^{(2)}(0)$ = 0.07 and exceptional stability over many hours of continuous operation. We also demonstrate a compact and robust confocal microscope design that enables scalable and practical deployment of SPS technologies. We anticipate that our work will spur new developments in quantum photonics and find applications in SPS-based QKD.

We employed cubic zirconia hemispheres with a diameter of 1 mm as the SILs for integration with hBN. hBN flakes were deposited onto the flat side of the SIL, and the sample was annealed to activate the SPSs. A custom-built portable confocal microscope was constructed to facilitate the measurements of the integrated sample at RT. The image of the microscope is shown in Figure 1 and the schematic of the microscope is shown as an inset. The design includes vibration isolating feet, removable sliding panels and a hinged lid to allow streamlined access to key components and interchanging of inputs and outputs. Additionally, a removable SPS mounting mechanism was used to enable the SPS exchange without affecting alignment or focus. The collection output can be configured to accept either a single-mode fibre or free-space collection. The entire confocal setup has a fingerprint of 500×500 mm² and weight load of ~ 10 kg, suitable for transport and space-based applications.

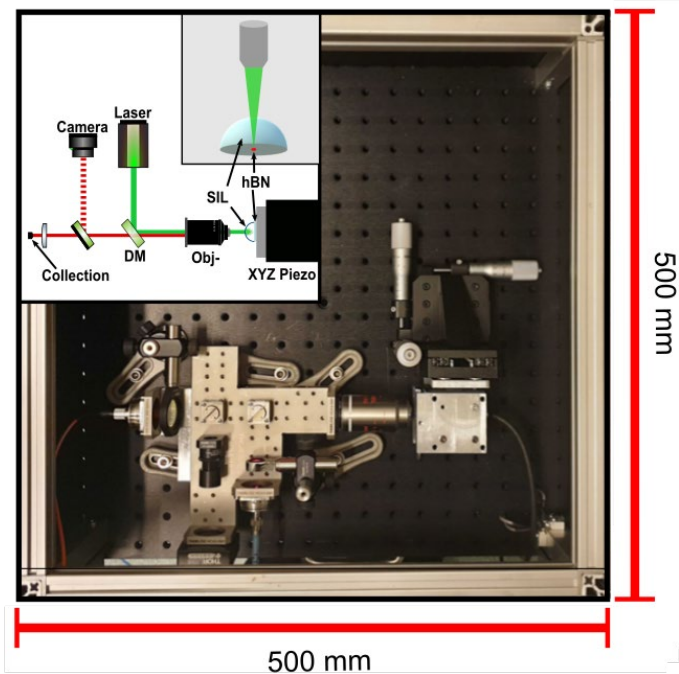

***Figure 1:*** *(a) Top-down photograph of the confocal setup in an isolated enclosure. Inset shows a schematic illustration of the key components in the confocal microscope, with an illustration of the hBN-SIL (solid-immersion lens) SPS in the top right corner. DM – dichroic mirror. Obj – objective. SIL – solid immersion lens.*

Figure 2a shows a characteristic hBN photoluminescence spectrum with a sharp zero phonon line (ZPL) at 728 nm, collected through the SIL. To prove the system is indeed an SPS, the collected light was directed into a Hanbury-Brown and Twiss (HBT) interferometer to perform a second-order correlation measurement, $g^2(\tau)$. For this measurement, pulsed excitation was used with a repetition rate of 40 MHz, removing the effects of timing jitter and reducing the background luminescence. The $g^2(\tau)$ measurement is shown in Figure 2b, with a $g^{(2)}(0)$ of 0.07, which is excellent for a RT SPS. The $g^{(2)}(0)$ was obtained by fitting the data with a Lorentzian function and the calculated areas under the peaks, taking the ratio of the integrated correlations at zero delay and $\tau$ time.

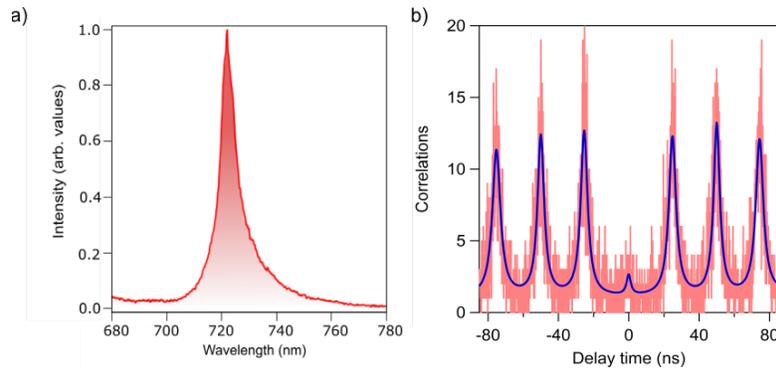

*Figure 2: (a) PL spectrum of a characteristic SPS. (b) Pulsed second-order autocorrelation measurement acquired at 40 MHz repetition rate, showing a $g^{(2)}(0)$ value of 0.07.*

The SIL used in figure 1 is intended to improve the collection efficiency, and hence the count rate and signal-to-noise ratio of the SPS. The employed cubic zirconia SIL acts as a high refractive-index material, in contrast to the hBN. Due to the resultant high index of refraction, the integrated system is expected to have an increased numerical aperture (NA), yielding a greater collection angle and thus an increased collection efficiency. To evaluate the effectiveness of the SIL, the SPS count rate was measured as a function of excitation power. Figure 3a shows a comparison of count rates through the SIL (blue trace) and the same pristine emitter without the SIL (red trace). The latter was collected by inverting the SIL. The experimental data was fitted to the three level system equation: $I = (I_{sat} P) / (P + P_{sat})$, where $I$, $(P)$ is the emission rate (excitation power) and $I_{sat}$, $P_{sat}$ is the emission and excitation power at saturation. The saturated count rate of the emitter with (without) the SIL is measured to be ~ $1.05 \times 10^7$ counts/s ($1.67 \times 10^6$ counts/s) as measured through free-space collection, representing over a six-fold increase. Additionally, the power needed to reach saturation was reduced to 73.1 µW when the SIL was employed compared to 223 µW for the refence measurement.

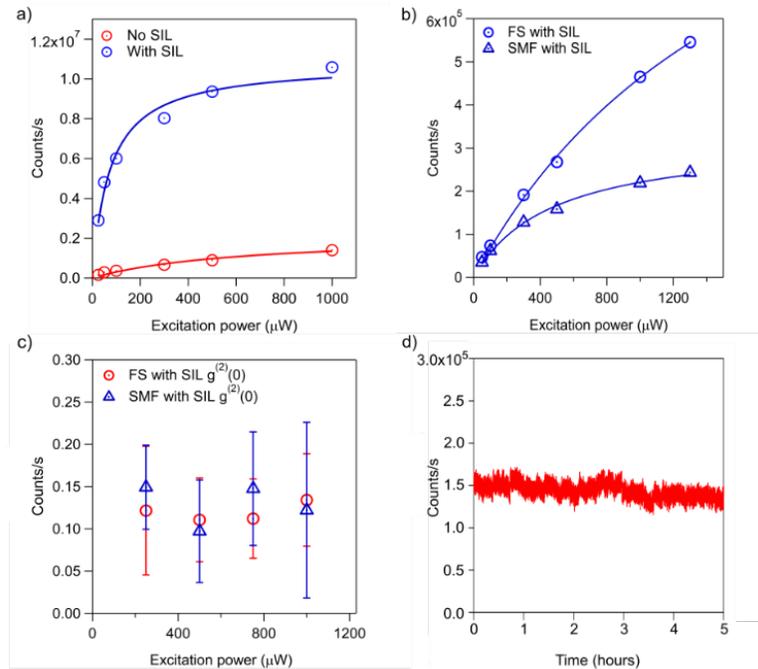

*Figure 3. (a) Power-dependent excitation measured with a CW laser source through the SIL (blue circles) and without the SIL (red circles). (b) Comparison of saturation count rates from a SIL-integrated SPS in free space (FS) (blue circles) and through a single mode fibre (SMF) (blue triangle). The data in*

*a, b are fit with the standard three level model. (c) Power-dependent g$^{(2)}$(τ) measured for the same emitter as in (b) at different laser powers, collected through an SMF and in free space. (d) Photostability demonstration of an SPS recorded over 5 hours of continuous acquisition.*

Given the importance of fibre communications, we also measured the emission rate through a single mode fibre (SMF), to determine how the increase in collection efficiency translates to the practical setup of a fibre-based system. A different characteristic hBN emitter (with a ZPL at 582 nm) integrated with the SIL was measured in free-space and using SMF collection, employing the same confocal setup shown in Figure 1. Figure 3b shows the result of this measurement—with observed count rates of ~ $5.46 \times 10^5$ counts/s and ~ $2.43 \times 10^5$ counts/s for free-space and SMF collection, respectively, translating to a coupling efficiency of ~ 44% through the SMF.

Next, we verify the stability and performance of the custom-built confocal microscope and enclosure. For practical purposes, the confocal system must perform to the same standard as typical lab-built confocal microscopes. Thus the following measurements were conducted to analyse signal-to-noise ratio and stability. Figure 3c shows power-dependent $g^{(2)}(\tau)$ measurements performed on the same SPS as seen in Figure 2b comparing $g^{(2)}(0)$ values between free-space and SMF collection. The performance between the free-space and SMF collection is observed to be of a similar value, showing the g$^{(2)}$(0) value to be within error bars for each respective excitation power. This is important to assess the performance and repeatability of our custom-built portable confocal microscope. Finally, Figure 3d shows the PL intensity from the SPS for an extensive duration of 5 hours. The observed count rates indicate the extreme photostability of the hBN source, absent of any photoblinking or large fluctuations in emission intensity, confirming also the mechanical robustness of the box. There was no physical realignment during the measurement.

Lastly, we provide a real-world evaluation of suitability of the SPS for practical QKD, considering the well-established BB84 protocol[24, 25]. We consider the security analysis in the finite-key length regime in [26] to obtain an estimate of the secret key rate for practical deployment. The achievable QKD key rate, $K$, (in bits per second) is given by $K=S_{\text{finite}}R_s$ where $S_{\text{finite}}$ is the secret key rate (in bits per pulse) extracted at the end of the QKD protocol per pulse and $R_s$ is the pulse repetition rate at the source. Note, the key rate $K$ obtained here assumes that the classical post-processing steps, i.e. the reconciliation and privacy amplification, are not the bottleneck of the overall QKD deployment. The assumption can be easily satisfied if state-of-the-art FPGA-based reconciliation and privacy amplification are adopted.

We note recent work[27] on the second-order correlation function, g$^{(2)}$(τ), that for some quantum state of light, the there is a non zero projection onto the single-photon Fock state. We refer to this projection as the single-photon projection. Specifically, we note a formal proof is given that a value of g$^{(2)}$(0) < 0.5 for a given state, implies that there is a nonzero single-photon projection in the state. However, the probability, $p$, of obtaining a single photon in a measurement of the photon number is completely arbitrary - it is not related to g$^{(2)}$(0). We therefore derive our key rate independent of the specifically measured g$^{(2)}$(0) value.

We define the total failure probability, ϵ, of the protocol as $\epsilon = \tilde{\epsilon} + \epsilon_{\text{PA}} + \epsilon_{\text{EC}} + \epsilon_{\text{PE}}$, where $\tilde{\epsilon}$ is the smoothing parameter for the smooth min-entropy calculation, $\epsilon_{\text{PA}}$ is the failure probability of privacy amplification, $\epsilon_{\text{EC}}$ is the failure probability of error-correction, and $\epsilon_{\text{PE}}$ is the probability that the true value of the Quantum Bit Error Rate (QBER) is out of the confidence. For a given ϵ, $S_{\text{finite}}$ is given by

$$S_{\text{finite}} = qA(1 - h\left(\frac{\tilde{\epsilon}(m)}{A}\right) - f_{\text{EC}}h(e) - \Delta(n)) \qquad (1)$$

Where $f_{EC}$ is the reconciliation efficiency, *q=0.5n/(n+m)* is the ratio of remaining bits after parameter estimation to the total number of bits obtained from the quantum measurement. Here, *n* is the number of bits that Alice and Bob hold for classical post-processing, and *m* is the number of bits randomly selected for parameter estimation. In Eq. 1, *h(x)* is the binary entropy function, the term, *A = ($P_{det}$ – $P_m$)/$P_{det}$*, is the correction term due to the multi-photon emission at the SPS, where $P_{det}$ is the probability of detecting at least one photon, and $P_m$ is the probability of multi-photon emission at the source. In Eq. (1), *e* is the true value of QBER, and $\tilde{\epsilon}(m)$ is the estimated QBER taking the statistical fluctuation of using those *m* bits for the QBER into account.

Based on a realistic value of $P_m$ = 0.07 from the previous measurements of our SIL-integrated SPSs, we investigate the dependence of *K* on the fibre length, utilising a fibre-based QKD system. The results are shown in Figure 4. The calculations show that key rates, *K*, approaching ~ $10^6$ bits/s can be achieved for 20 MHz repetition rates (exceeding that for faster triggering). The scale at which non-zero key rate are found (up to ~ 8 km radius), are sufficient to cover the core metropolitan area of a typical city.

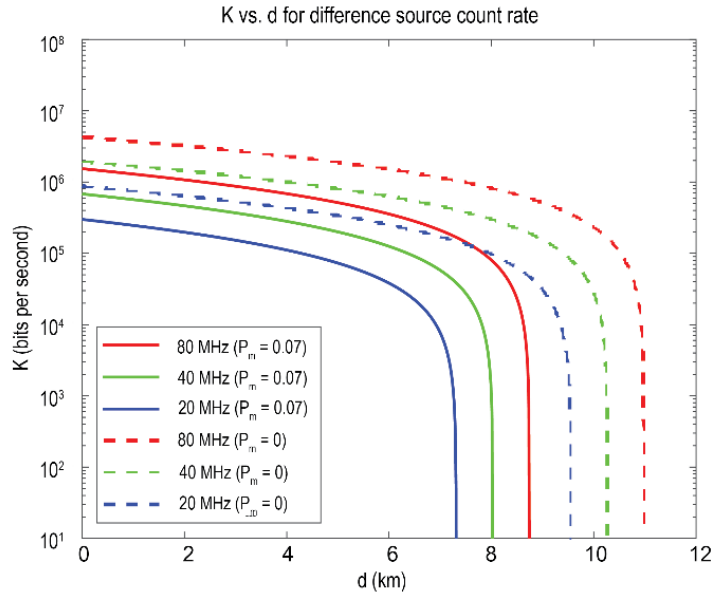

**Figure 4.** K vs. d for different Rs. We adopt the following setup to obtain this figure: The fibre attenuation factor for 720nm fibre optic, α = 3.5 db/km, e = 2%, ϵ = $10^{-10}$, n = $10^6$, $f_{EC}$ = 1.1, and *m = 0.5n*.

To summarise, we show that an integrated hBN-SIL SPS can provide single-photon rates of $10^7$ Hz, and can possess a high purity of $g^{(2)}(0)$ = 0.07. The sources are bright, photostable over a prolonged duration and can be incorporated into a portable, lightweight setup. We also show the potential of employing such a source under realistic conditions and show that high secured QKD rates approaching ~ MHz are feasible over several km. We expect the presented hBN-SIL system will accelerate the development of practical QKDs systems and will find place in real-world quantum photonic applications.

**Funding**
We thank the NSW Defence Innovation Network, the NSW State Government and the Next Generation Technologies Fund for financial support of this project funded by the DIN Strategic Investment Initiative.

**Acknowledgments**
The authors thank OptoFab node of the ANFF for access to facilities

**Disclosure**
The authors declare no conflicts of interest

**REFERENCES**
1. Senellart, P., G. Solomon, and A. White, *High-performance semiconductor quantum-dot single-photon sources.* Nature Nanotechnology, 2017. **12**: p. 1026.
2. Aharonovich, I., D. Englund, and M. Toth, *Solid-state single-photon emitters.* Nature Photonics, 2016. **10**(10): p. 631-641.
3. Beveratos, A., R. Brouri, T. Gacoin, A. Villing, J.P. Poizat, and P. Grangier, *Single photon quantum cryptography.* Physical Review Letters, 2002. **89**(18): p. 187901.
4. Kupko, T., M. von Helversen, L. Rickert, J.-H. Schulze, A. Strittmatter, M. Gschrey, S. Rodt, S. Reitzenstein, and T. Heindel, *Tools for the performance optimization of single-photon quantum key distribution.* npj Quantum Information, 2020. **6**(1): p. 29.
5. Wang, J., F. Sciarrino, A. Laing, and M.G. Thompson, *Integrated photonic quantum technologies.* Nature Photonics, 2020. **14**(5): p. 273-284.
6. Lo, H.-K., M. Curty, and K. Tamaki, *Secure quantum key distribution.* nature Photonics, 2014. **8**(8): p. 595-604.
7. Takemoto, K., Y. Nambu, T. Miyazawa, Y. Sakuma, T. Yamamoto, S. Yorozu, and Y. Arakawa, *Quantum key distribution over 120 km using ultrahigh purity single-photon source and superconducting single-photon detectors.* 2015. **5**: p. 14383.
8. Leifgen, M., T. Schröder, F. Gädeke, R. Riemann, V. Métillon, E. Neu, C. Hepp, C. Arend, C. Becher, K. Lauritsen, and O. Benson, *Evaluation of nitrogen- and silicon-vacancy defect centres as single photon sources in quantum key distribution.* New Journal of Physics, 2014. **16**(2): p. 023021.
9. Tran, T.T., K. Bray, M.J. Ford, M. Toth, and I. Aharonovich, *Quantum emission from hexagonal boron nitride monolayers.* Nat Nanotechnol, 2016. **11**(1): p. 37-41.
10. Grosso, G., H. Moon, B. Lienhard, S. Ali, D.K. Efetov, M.M. Furchi, P. Jarillo-Herrero, M.J. Ford, I. Aharonovich, and D. Englund, *Tunable and high-purity room temperature single-photon emission from atomic defects in hexagonal boron nitride.* Nature Communications, 2017. **8**(1): p. 705.
11. Vogl, T., R. Lecamwasam, B.C. Buchler, Y. Lu, and P.K. Lam, *Compact Cavity-Enhanced Single-Photon Generation with Hexagonal Boron Nitride.* ACS Photonics, 2019. **6**(8): p. 1955-1962.
12. Li, C., Z.-Q. Xu, N. Mendelson, M. Kianinia, M. Toth, and I. Aharonovich, *Purification of single-photon emission from hBN using post-processing treatments.* Nanophotonics, 2019. **8**(11): p. 2049-2055.
13. Exarhos, A.L., D.A. Hopper, R.R. Grote, A. Alkauskas, and L.C. Bassett, *Optical Signatures of Quantum Emitters in Suspended Hexagonal Boron Nitride.* ACS Nano, 2017. **11**(3): p. 3328-3336.
14. Jungwirth, N.R., B. Calderon, Y. Ji, M.G. Spencer, M.E. Flatt, and G.D. Fuchs, *Temperature Dependence of Wavelength Selectable Zero-Phonon Emission from Single Defects in Hexagonal Boron Nitride.* Nano Letters, 2016. **16**(10): p. 6052-6057.
15. Tran, T.T., K. Bray, M.J. Ford, M. Toth, and I. Aharonovich, *Quantum emission from hexagonal boron nitride monolayers.* Nature Nanotechnology, 2016. **11**(1): p. 37-41.
16. Bogdanov, S.I., M.Y. Shalaginov, A.S. Lagutchev, C.-C. Chiang, D. Shah, A.S. Baburin, I.A. Ryzhikov, I.A. Rodionov, A.V. Kildishev, A. Boltasseva, and V.M. Shalaev, *Ultrabright Room-Temperature Sub-Nanosecond Emission from Single Nitrogen-Vacancy Centers Coupled to Nanopatch Antennas.* Nano Letters, 2018. **18**(8): p. 4837-4844.
17. Hoang, T.B., G.M. Akselrod, C. Argyropoulos, J. Huang, D.R. Smith, and M.H. Mikkelsen, *Ultrafast spontaneous emission source using plasmonic nanoantennas.* Nature Communications, 2015. **6**: p. 7788-7788.

18. Nguyen, M., S. Kim, T.T. Tran, Z.Q. Xu, M. Kianinia, M. Toth, and I. Aharonovich, *Nanoassembly of quantum emitters in hexagonal boron nitride and gold nanospheres.* Nanoscale, 2018. **10**(5): p. 2267-2274.
19. Sapienza, L., M. Davanço, A. Badolato, and K. Srinivasan, *Nanoscale optical positioning of single quantum dots for bright and pure single-photon emission.* Nature Communications, 2015. **6**(1): p. 7833.
20. Moczała-Dusanowska, M., Ł. Dusanowski, O. Iff, T. Huber, S. Kuhn, T. Czyszanowski, C. Schneider, and S. Höfling, *Strain-Tunable Single-Photon Source Based on a Circular Bragg Grating Cavity with Embedded Quantum Dots.* ACS Photonics, 2020. **7**(12): p. 3474-3480.
21. Vamivakas, A.N., R.D. Younger, B.B. Goldberg, A.K. Swan, M.S. Ünlü, E.R. Behringer, and S.B. Ippolito, *A case study for optics: The solid immersion microscope.* American Journal of Physics, 2008. **76**(8): p. 758-768.
22. Marseglia, L., J.P. Hadden, A.C. Stanley-Clarke, J.P. Harrison, B. Patton, Y.-L.D. Ho, B. Naydenov, F. Jelezko, J. Meijer, P.R. Dolan, J.M. Smith, J.G. Rarity, and J.L. O'Brien, *Nanofabricated solid immersion lenses registered to single emitters in diamond.* applied physics letters, 2011. **98**(13): p. 133107.
23. Hwang, J., M. Pototschnig, R. Lettow, G. Zumofen, A. Renn, S. Gotzinger, and V. Sandoghdar, *A single-molecule optical transistor.* Nature, 2009. **460**(7251): p. 76-80.
24. Bennett, C.H. and G. Brassard. *Quantum cryptography: Public key distribution and coin tossing*. 1984. New York.
25. Lütkenhaus, N., *Security against individual attacks for realistic quantum key distribution.* Physical Review A, 1999. **61**.
26. Chaiwongkhot, P., S. Hosseini, A. Ahmadi, B. Higgins, D. Dalacu, P. Poole, R. Williams, M. Reimer, and T. Jennewein, *Enhancing secure key rates of satellite QKD using a quantum dot single-photon source*. 2020.
27. Grünwald, P., *Effective second-order correlation function and single-photon detection.* New Journal of Physics, 2019. **21**(9): p. 093003.